\documentclass{PoS}
\usepackage{subfig}
\usepackage{lineno} % MARCOS

\title{AMON: TeV Gamma and Neutrino Coincidence Alerts from HAWC and IceCube subthreshold data}

\ShortTitle{TeV Gamma and Neutrino Coincidence Alerts}

\author{
{\bfseries Corresponding authors:}
\speaker{Hugo Alberto Ayala Solares}$^{1}$\\
The HAWC Collaboration\footnote{For collaboration list, see PoS(ICRC2019) 1177.}\\
{ \itshape \href{https://www.hawc-observatory.org/collaboration/icrc2019.php}{https://www.hawc-observatory.org/collaboration/icrc2019.php}}\\
The IceCube Collaboration\footnote{For collaboration list, see PoS(ICRC2019) 1177.}\\
{\itshape \href{http://icecube.wisc.edu/collaboration/authors/icrc19_icecube}{http://icecube.wisc.edu/collaboration/authors/icrc19\_icecube}}\\
The AMON Project\\
{$^{1}$ \itshape The Pennsylvania State University}\\
E-mail: \email{hgayala@psu.edu}
}

\abstract{
The era of multimessenger astrophysics has arrived with the simultaneous
operation of large cosmic-ray, gamma-ray, neutrino, and gravitational-wave
observatories. In just the past two years, an electromagnetic (EM) counterpart was detected for a gravitational wave event, and evidence for an EM counterpart of high energy neutrinos has been identified. These measurements
have had a major impact on our view of the non-thermal universe, but
understanding cosmic accelerators require a substantial increase in the number
of multimessenger observations. The Astrophysical Multimessenger Observatory
Network (AMON) is designed for high-statistics searches of sub-threshold
transient alerts from gamma-ray and neutrino detectors. Within AMON, we have
implemented a joint-likelihood analysis of TeV gamma-ray measurements from
the High Altitude Water Cherenkov (HAWC) Observatory and neutrinos from the
IceCube Neutrino Observatory. AMON is ready to produce real-time coincidence
alerts using HAWC ``hotspots'' and IceCube astrophysical neutrino events. These
alerts will be distributed to AMON follow-up partners with a median anticipated
delay of six hours, which corresponds to a full transit in the field of view of
HAWC. The alerts will have an angular resolution of ${\sim}0.2^{\circ}$, making them well-
suited for deep electromagnetic follow-up observations.\\

}

\FullConference{36th International Cosmic Ray Conference -ICRC2019-\\
		July 24th - August 1st, 2019\\
		Madison, WI, U.S.A.}

\begin{document}

\section{Introduction}\label{sec:info}
The realm of multimessenger astrophysics had a new rebirth after the coincident detection of gravitational waves and electromagnetic radiation~\cite{bns}, as well as the identification of evidence for coincident detection of neutrinos and gamma-ray flare from a blazar~\cite{icneutrino}. 
In the present proceedings we will focus on a coincidence analysis combining information from the HAWC gamma-ray observatory~\cite{hawc} and the IceCube~\cite{icecube} neutrino observatory. For this we will use the infrastructure of the Astrophysical Multimessenger Observatory Network (AMON).

AMON has been created with the purpose of facilitating the interaction of the different observatories, creating a framework capable of doing analyses with different datasets, and alerting the astrophysical community of any interesting events worthy of subsequent follow-up~\cite{amonI,amonII}\footnote{AMON webpage\cite{amonIII}}.

The analyisis presented here is developed under the AMON framework. The purpose of the this analysis is to search for transient or variable sources that can produce both gamma rays and neutrinos, specially sources like blazars or radio galaxies~\cite{gammanuemissionII}. 

\section{HAWC and IceCube $\gamma + \nu$ alerts}\label{sec:data}

%\subsection{HAWC Data}
The HAWC observatory is a high-energy gamma-ray detector, located in central Mexico. 
The complete detector has been in operation since March 2015. 
HAWC has a large field of view, covering two thirds of the sky daily with a high-duty cycle in the declination range from $-26^{\circ}$ to $64^{\circ}$. HAWC is sensitive to gamma rays in the energy range between ${\sim}100$~GeV and ${\sim}100$~TeV. It has an angular resolution of $0.1^{\circ}{-}1.0^{\circ}$.
Among HAWC's many successes are: the detection of TeV jets from SS433 and discovery of a new class of halos, named TeV-Halos, such as Geminga~\cite{geminga}.

As an input to our search we use any significant excess, or ``hotspot'', locations in the sky with a cluster of events above the estimated cosmic-ray background level, identified during one full transit of that sky location above the detector.
The main ``hotspot'' parameters AMON receives are: the sky position and its uncertainty, significance value ($>2.75\sigma$) and the start and stop times of the transit. The threshold in the significance was defined by the HAWC collaboration to send sub-threshold events. The current rate of these ``hotspot'' received by AMON is $\sim 800$ per day.

%\subsection{IceCube Data}
The IceCube observatory is  a detector of high-energy neutrinos located at the South Pole.  The detector construction started in 2005 and finished in 2010. 
IceCube can search for neutrinos from the whole sky, though it is more sensitive to searches from the northern celestial hemisphere since the Earth helps reduce the measured atmospheric muon background in IceCube. This is an advantage for the analysis since HAWC's most sensitive region corresponds to the northern hemisphere. 
IceCube is sensitive to energies that can reach up to EeV near the horizon ($\delta=0^{\circ}$).
The angular resolution depends on the topology of signal events inside the detector. Two main topologies are observed: track events and cascade events. Track events are mostly formed by charged-current muon-neutrino interactions. This tracks can have a length of $\mathcal{O}$(km). The track events have a median angular resolution of ${\sim}0.4^{\circ}$ above 100 TeV. 
Cascade events are produced by the other types of neutrinos, or neutral-current interactions of muon neutrinos. They have better energy resolution compared to tracks since the signal of the events are completely contained inside the detector. Their angular resolution, however, is ${>}10^{\circ}$.

%In September 2017, evidence was found that the IceCube event 170922A was coincident with enhanced gamma-ray emission from the blazar TXS 0506+056~\cite{icneutrino}. %The Fermi-LAT telescope had observed previously gamma-ray emission from this blazar. This gamma-ray emisison by itself would have not be interesting, but having found that it occurred 7 seconds after the neutrino emission made it one of the most followed-up objects during 2017-2018. 

The IceCube data sent to AMON consists of single through-going track events. The parameters consists of: sky position and its uncertainty, time of the event, false positive rate density (number of events in the data in a zenith angle bin, above an energy threshold, per livetime per solid angle) and signal acceptance. The current rate of the events received by AMON is $\sim 650$ per day.

\section{Method}\label{sec:analysis}

The coincidence analysis is defined by two criteria. First is a temporal one, where we look for neutrinos inside the transit time of the HAWC ``hotspot''. Second, we select neutrinos that are within a radius of $3.5^{\circ}$ from the HAWC ``hotspot'' localization. 
After the neutrino events have passed the selection criteria, we calculate a ranking statistic to select the most interesting coincident events. 
This ranking statistic is based on Fisher's method, where we combine all the information that we have from the events. It is defined as:
\begin{equation}\label{eq:rank1}
\chi^2_{6+2n_{\nu}}=-2 \ln[p_{_{\lambda}}p_{_{HAWC}}p_{_{cluster}}\prod_i^{\nu}p_{_{IC,i}}],
\end{equation}
where $p_{_{\lambda}}$ quantifies the overlap of the spatial uncertainties of the events; $p_{_{HAWC}}$ is the probability of the HAWC event being compatible with a background fluctuation; $p_{_{cluster}}$ is the probability of seeing more than one neutrino from background in the HAWC transit period; and $p_{_{IC,i}}$ is the probability of the IceCube event been from background. 
The only p-value that we have to calculate within the AMON framework is the $p_{_{\lambda}}$, which comes from maximizing a likelihood calculation that measures how much the position of the HAWC event and the IceCube events overlap with each other. This is calculated as 

\begin{equation}
\lambda(\vec{x}) = \sum_{i=1}^N  (ln(S_i(\vec{x})) - ln(B_i))
\end{equation}

where $S$ corresponds to the uncertainties of the events, assuming Gaussian distributions on the sphere, and $B$ is the spatial background distribution from each detector at the position of the events. This likelihood is maximized by finding the best position of the coincidence $\vec{x}$.  A higher $\lambda$ value means the uncertainties of the events overlap more. This translates into a smaller $p_{_{\lambda}}$.

Due to the fact that we can have more than one neutrino in the time window, this affects the degrees of freedom of equation \ref{eq:rank1}. Considering this, we transform the $\chi^2$ to a p-value, with the corresponding number of degrees of freedom, and then calculate the negative logarithm of this quantity. This is represented as
\begin{equation}\label{eq:rank2}
\chi^{2'} = -\log p(>\chi^{2}_{6+2n_{\nu}}),
\end{equation}
which is the value that we used to rank the coincidences. 

\section{Calibration of the False Alarm Rate}\label{sec:simulations}
\begin{figure}
    \centering
    \includegraphics[scale=0.5]{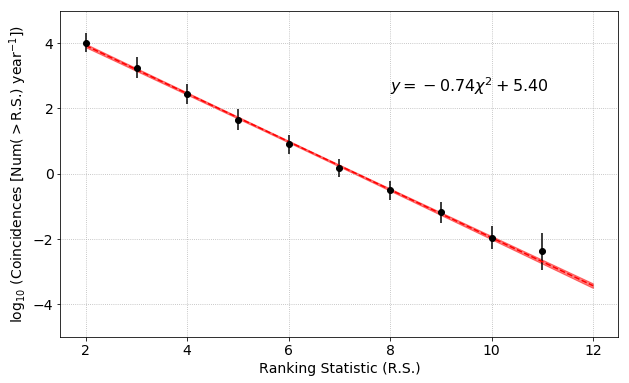}
    \caption{False alarm rate as a function of the ranking statistic obtained from the scrambled datasets. This is the main function that will be used to select alerts that will be sent to the Galactic Coordinates Network~\cite{gcn}. A false alarm rate of 1 per year is obtained with a ranking statistic value of 7.3.}
    \label{fig:rates}
\end{figure}

We used the previous algorithm on scrambled datasets based on 2 years of data from both observatories. The scrambling process was applied several times for testing the analysis. This corresponds to a $\sim$729 years of data. 
The scrambling consisted on randomly permuting the right ascension information of the events. Parameters that are declination dependent were kept together with their respective declination (e.g. false positive rate density in IceCube, and transit time in HAWC). 
The main purpose of this exercise was to obtain the false alarm rate of this analysis, which is shown in Fig. \ref{fig:rates}

\section{Unblinding Results}\label{sec:results}

\begin{figure}
    \centering
    \includegraphics[scale=0.5]{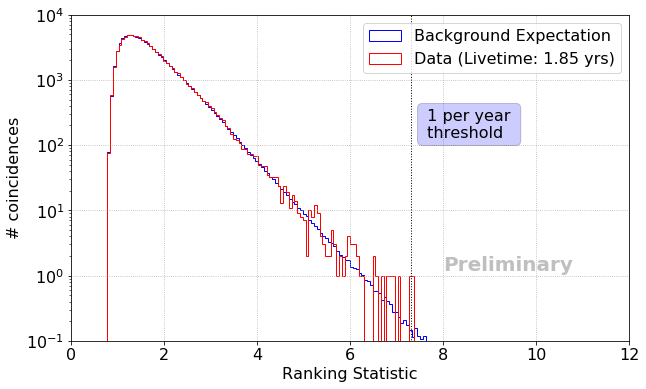}
    \caption{Ranking statistics distribution of the analysis. Blue: background expectation obtained from the scrambled datasets and normalized to the number of coincidences observed in the unblinded dataset. Red: result from the unblinded analysis.}
    \label{fig:coincdist}
\end{figure}

After obtaining the false alarm rate (FAR) for the analysis, we unblinded two years of data (with a livetime of 1.85 years). Figure \ref{fig:coincdist} shows the distribution of the ranking statistic of the unblinded two years of data compared to the distribution obtained with the scrambled datasets (normalized to the total counts from the unblinded set). 

\begin{figure}%
	\centering
	\subfloat[Coincidence with a false alarm rate of 0.99 per year]{{ \includegraphics[scale=0.45,clip]{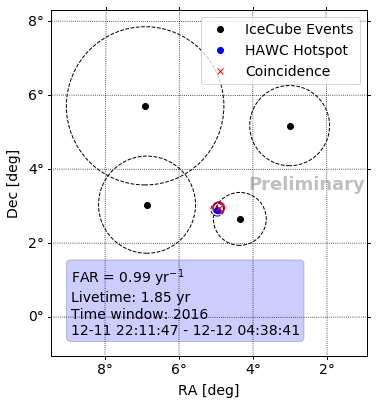} }}%
	\qquad
	\subfloat[Coincidence with a false alarm rate of 1.1 per year]{{ \includegraphics[scale=0.45,clip]{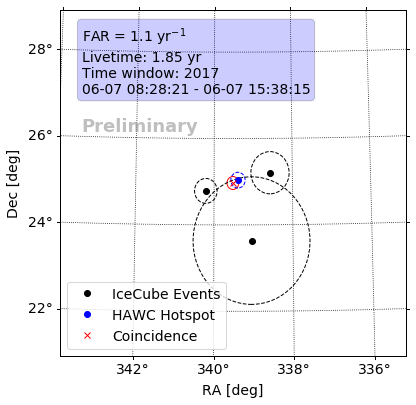} }}%
	\caption{Skymaps of the coincidences found with the analysis. The size of the circles correspond to the angular uncertainties of the positions ($1\sigma$ for a 2D Gaussian). Size of the uncertainty radius of the coincidence is ${\sim}0.1^{\circ}$ after combining the uncertainties.}
	\label{fig:skymaps}
\end{figure}

We found one coincidence event that passed the threshold of 7.3 in ranking statistic that corresponds to a FAR of one coincidence per year, with a value of 7.34. Another coincidence, which was close to the threshold, has a value of 7.27. The skymap figures of the two coincident events with the highest ranking statistic values in this datasets are shown in Fig. \ref{fig:skymaps}

\section{Conclusions and Outlook}\label{sec:concl}
We developed a method to search for coincidences of sub-threshold data from the HAWC and the IceCube observatories. The method was tested on archival data during the years 2016 and 2017. 
We found one event that crosses the false alarm rate threshold of one per year.

This method will serve as a base for a real-time search for multimessenger sources of neutrinos, as well as to search for the sources of ultra-high energy cosmic rays. 
The search would consist of sending alerts that have a false alarm rate of one per year to follow-up observatories. 

\acknowledgments

\small{
AMON: This research or portions of this research were conducted with Advanced CyberInfrastructure computational resources provided by the Institute for CyberScience (\href{https://ics.psu.edu}{https://ics.psu.edu}) at the Pennsylvania State University . This material is based upon work supported by the National Science Foundation under Grants PHY-1708146 and PHY-1806854 and by the Institute for Gravitation and the Cosmos of the Pennsylvania State University. Any opinions, findings, and conclusions or recommendations expressed in this material are those of the author(s) and do not necessarily reflect the views of the National Science Foundation.

HAWC: We acknowledge the support from: the US National Science Foundation (NSF) the US Department of Energy Office of High-Energy Physics; 
the Laboratory Directed Research and Development (LDRD) program of Los Alamos National Laboratory; 
Consejo Nacional de Ciencia y Tecnolog\'{\i}a (CONACyT), M{\'e}xico (grants 271051, 232656, 260378, 179588, 254964, 271737, 258865, 243290, 132197, 281653)(C{\'a}tedras 873, 1563, 341), Laboratorio Nacional HAWC de rayos gamma; 
L'OREAL Fellowship for Women in Science 2014; 
Red HAWC, M{\'e}xico; 
DGAPA-UNAM (grants AG100317, IN111315, IN111716-3, IA102715, IN109916, IA102019, IN112218); 
VIEP-BUAP; 
PIFI 2012, 2013, PROFOCIE 2014, 2015; 
the University of Wisconsin Alumni Research Foundation; 
the Institute of Geophysics, Planetary Physics, and Signatures at Los Alamos National Laboratory; 
Polish Science Centre grant DEC-2014/13/B/ST9/945, DEC-2017/27/B/ST9/02272; 
Coordinaci{\'o}n de la Investigaci{\'o}n Cient\'{\i}fica de la Universidad Michoacana; Royal Society - Newton Advanced Fellowship 180385. Thanks to Scott Delay, Luciano D\'{\i}az and Eduardo Murrieta for technical support.

IceCube: We acknowledge the support from the following agencies: U.S. National Science Foundation-Office of Polar Programs, U.S. National Science Foundation-Physics Division, University of Wisconsin Alumni Research Foundation, the Grid Laboratory Of Wisconsin (GLOW) grid infrastructure at the University of Wisconsin - Madison, the Open Science Grid (OSG) grid infrastructure; U.S. Department of Energy, and National Energy Research Scientific Computing Center, the Louisiana Optical Network Initiative (LONI) grid computing resources; Natural Sciences and Engineering Research Council of Canada, WestGrid and Compute/Calcul Canada; Swedish Research Council, Swedish Polar Research Secretariat, Swedish National Infrastructure for Computing (SNIC), and Knut and Alice Wallenberg Foundation, Sweden; German Ministry for Education and Research (BMBF), Deutsche Forschungsgemeinschaft (DFG), Helmholtz Alliance for Astroparticle Physics (HAP), Initiative and Networking Fund of the Helmholtz Association, Germany; Fund for Scientific Research (FNRS-FWO), FWO Odysseus programme, Flanders Institute to encourage scientific and technological research in industry (IWT), Belgian Federal Science Policy Office (Belspo); Marsden Fund, New Zealand; Australian Research Council; Japan Society for Promotion of Science (JSPS); the Swiss National Science Foundation (SNSF), Switzerland; National Research Foundation of Korea (NRF); Villum Fonden, Danish National Research Foundation (DNRF), Denmark
}

% Set up the bibliography using BibTeX.
% Get references from inspirehep.net or NASA/ADS and put them in references.bib.
\bibliographystyle{ICRC}
%\bibliography{references}
\begingroup
\def\bibfont{\footnotesize}

\endgroup

\end{document}